\documentclass{article}

\usepackage{color}
\usepackage{graphicx}
\usepackage{amsmath}

\newcommand{\exam}{\begin{flushleft}\color{blue}\footnotesize\begin{tabular}{lll}}
\newcommand{\maxe}{\end{tabular}\color{black}\end{flushleft}}

\begin{document}
\color{black}
\title{A Note on Exhaustive State Space Search for Efficient Code Generation}
\author{Aart J.C. Bik \\ \texttt{aartbik@gmail.com}}
\date{}
\maketitle


\abstract{
This note explores state space search to find efficient instruction sequences that perform particular data manipulations. Once found, the instruction sequences are hard-wired in the code generator that needs these data manipulations. Since state space is only searched while developing the compiler, search time is not at a premium, which allows exhaustively searching for the best possible instruction sequences.}

\section{Introduction}

Compilers must often emit instruction sequences that accomplish particular data manipulations in the generated code. For example, a compiler may have to generate instructions that swap the contents of two scalar registers prior to an instruction with strict constraints on its register operands. Or, as another example, a compiler may have to emit instructions that broadcast the value in a scalar register to all elements of a vector register in the prologue of a vector loop. Using an efficient instruction sequence for each desired data manipulation reduces the runtime of any application that executes these data manipulations frequently.

Clearly, an optimizing compiler could try to find efficient instruction sequences during actual code generation. Although this possibly provides additional context for optimization, the major drawback of this approach is that search time directly contributes to compile-time during AOT compilation or, worse, runtime during JIT compilation. Alternatively, efficient instruction sequences for desired data manipulations could be searched for earlier, i.e. while the compiler is still being developed. Although this may provide less opportunities to exploit code context, search time is not at a premium in this approach, and an exhaustive state space search to find the best possible instruction sequences becomes possible. Once found, instruction sequences are hard-wired in the code generator and become at the immediate disposal of the compiler.

In this note, we explore using Prolog~\cite{colmerauer93} for such a state space search. To keep the presentation brief, we focus on finding efficient Intel SSE instruction sequences for a few simple SIMD data manipulations. However, the presented ideas easily generalize to other instructions sets and code generation problems.  

\section{State Space Search}

Finding an efficient instruction sequence to accomplish a particular data manipulation can be expressed as a {\bf state space search} problem~\cite{luger,nilsson}, with the original contents of memory and registers as \emph{start state}, the machine instructions as \emph{transitions} from one state to another state, and the desired contents of memory and registers as \emph{goal state}. A path from the start state to the goal state provides a solution to the problem. The best solution is given by the shortest path, i.e. the path with minimal length if all transitions have the same cost, or the path with minimal total weight if different transitions have varying costs, such as different cycle counts for the instructions.

\subsection{State Space}

As stated before, for the sake of brevity, we focus on finding efficient Intel SSE instruction sequences for a few simple SIMD data manipulations. Furthermore, to keep the state space size manageable, we focus on just a subset of the SIMD state, data types, and instructions, abstracting away from details related to general-purpose registers and instructions, state flags, memory operands, etc. In this simplified view, the SIMD state is fully defined by eight \texttt{xmm}-registers, represented in Prolog as a list with eight variables (variables start with an uppercase letter).

\exam
\verb"[ XMM0, XMM1, XMM2, XMM3, XMM4, XMM5, XMM6, XMM7 ]"
\maxe

Here, each variable can be bound to a Prolog term that represent particular contents, such as a list \texttt{[17.5, 11.9]} to denote a packed double-precision floating-point data type with the given numerical values, an atom \texttt{xmm0} (atoms start with a lowercase letter) to denote a particular but otherwise non-exploitable value, or the anonymous variable \texttt{\_} to denote any term. For example, the following list denotes a SIMD state in which registers 0, 3, and 7 contain packed data types with the given numerical values, registers 1 and 2 have particular but different contents that are not subject to further inspection, and all other registers are undefined.

\exam
\verb"[ [0,0,0,0], xmm1, xmm2, [8,7,6,5,4,3,2,1], _, _, _, [1.0, 2.5] ]"
\maxe

In the remainder of the paper, we will just consider \emph{packed dwords} represented by 4-elements lists, with the convention that the higher to lower packed elements appear left-to-right in the list.

\subsection{Transitions}

Each instruction transforms a SIMD state into another SIMD state. These transitions are modeled by a set of Prolog rules for each Intel SSE instruction in our simplified model. Each rule is this set will have the form
\exam
\verb"i(instruction, op1, op2, S, T)."
\maxe
to indicate that applying \texttt{instruction} to the given operands transitions from state \texttt{S} to state \texttt{T}. For example, the change in SIMD state by executing instruction
\exam
\verb"pxor xmm0, xmm0"
\maxe
is modeled by rule show below, which specifies that any contents of register \texttt{xmm0} (the anonymous variable \texttt{\_}) is zeroed out (the list \texttt{[0,0,0,0]}) while the contents of all registers remain unaffected.
\exam
\verb"i(pxor, xmm0, xmm0," \\
\verb"  [         _, XMM1, XMM2, XMM3, XMM4, XMM5, XMM6, XMM7 ]," \\
\verb"  [ [0,0,0,0], XMM1, XMM2, XMM3, XMM4, XMM5, XMM6, XMM7 ])."
\maxe

Similarly, using Intel syntax, where the destination register appears first, the change in SIMD state after
\exam
\verb"paddd xmm1, xmm7"
\maxe
is modeled by the following rule, which adds the packed integral elements of one register to the packed integral elements of another register.
\exam
\verb"i(paddd, xmm1, xmm7," \\
\verb"  [ XMM0, [A,B,C,D], XMM2, XMM3, XMM4, XMM5, XMM6, [E,F,G,H] ]," \\
\verb"  [ XMM0, [A+E,B+F,C+G,D+H], XMM2, XMM3, XMM4, XMM5, XMM6, [E,F,G,H] ]," \\
\maxe

Although this allows Prolog to reason about the instruction symbolically, sometimes we are also interested in evaluating the values using integral arithmetic. To that end, the following rule is added as well, which evaluates expressions in which all values are integers (such rules could be refined further to allow for partial evaluation).
\exam
\verb"i(paddd, xmm1, xmm7," \\
\verb" [ XMM0, [A,B,C,D], XMM2, XMM3, XMM4, XMM5, XMM6, [E,F,G,H] ]," \\
\verb" [ XMM0, [P,Q,R,S], XMM2, XMM3, XMM4, XMM5, XMM6, [E,F,G,H] ]) :-" \\
\verb" integer(A), integer(E), P is A+E, integer(B), integer(F), Q is B+F," \\
\verb" integer(C), integer(G), R is C+G, integer(D), integer(H), S is D+H."
\maxe

Similar rules are added for all other arithmetic, logical, comparison, and conversion instructions, and for all combinations of register pairs. A data shuffling instruction such as
\exam
\verb"punpckldq, xmm0, xmm3"
\maxe
is modeled as shown below.
\exam
\verb"i(punpckldq, xmm0, xmm3," \\
\verb" [ [_,_,A,B], XMM1, XMM2, [X,Y,C,D], XMM4, XMM5, XMM6, XMM7 ]," \\
\verb" [ [C,A,D,B], XMM1, XMM2, [X,Y,C,D], XMM4, XMM5, XMM6, XMM7 ])."
\maxe

A shift instruction like
\exam
\verb"psrldq xmm2, 4"
\maxe
is modeled by the rule below.
\exam
\verb"i(psrldq, xmm2, 4," \\
\verb"  [ XMM0, XMM1, [A,B,C,_], XMM3, XMM4, XMM5, XMM6, XMM7 ]," \\
\verb"  [ XMM0, XMM1, [0,A,B,C], XMM3, XMM4, XMM5, XMM6, XMM7 ])."
\maxe

The SIMD state change after the data movement instruction
\exam
\verb"movd xmm4, I"
\maxe
is modeled with this rule.
\exam
\verb"i(movd, xmm4, I," \\
\verb" [ XMM0, XMM1, XMM2, XMM3,         _, XMM5, XMM6, XMM7 ]," \\
\verb" [ XMM0, XMM1, XMM2, XMM3, [0,0,0,I], XMM5, XMM6, XMM7 ])."
\maxe

Obviously, writing all these Prolog rules by hand would be too tedious and error-prone. Instead, a utility should be used to generate all rules automatically, preferably directly from an instruction set description in an electronic format. A complete and accurate rule set will obviously yield the best results.

\subsection{Search}

Given a complete Prolog rule set that model the SIMD state transitions of all instructions, we need a search mechanism to find a path in the state space from the \emph{start state} to the \emph{goal state}. This search mechanism is also expressed with Prolog rules.

A first reasonable attempt is shown below (we will refine these rules slightly later). The two rules states that any state \texttt{S} transitions into itself for an empty instruction sequence, or otherwise breaks down into the transition of a single instruction from state \texttt{S} to state \texttt{U} followed by the transition from state \texttt{U} to state \texttt{T} of an subsequent instruction sequence \texttt{J}. The list of 3-arity \texttt{i} predicates built by these rules ultimately indicate an instruction sequence that transitions state \texttt{S} to state \texttt{T}.   

\exam
\verb"s(S, [], S)." \\
\verb"s(S, [i(I,R1,R2)|J], T) :- i(I, R1, R2, S, U), s(U, J, T)."
\maxe

Now suppose we are interested in finding the best way to zeroing the contents of register \texttt{xmm0}. It may be tempting to express that particular state space search problem with the following Prolog query. 

\exam
\verb"s(S, I, [[0,0,0,0] | _ ])."
\maxe
However, this query returns the following first solution, with as interpretation that the shortest way of resetting register \texttt{xmm0} to zero is by executing no instructions at all (empty list \texttt{I}) but instead starting with all zeroes in that register (initial state \texttt{S}). Although correct, this is obviously not what we were searching for.
\exam
\verb"I = []" \\
\verb"S = [[0,0,0,0]|_]"
\maxe

As a side note, this mistake can demonstrate a potential danger of using anonymous variables. The almost identical query
\exam
\verb"s(_, I, [[0,0,0,0] | _ ])." 
\maxe
would have given the solution
\exam
\verb"I = []"
\maxe
as well, but without even listing bindings for the two anonymous variables, obscuring the fact that the initial state was bound to a state with the first register already zeroed out. So each anonymous variable really denotes any \emph{suitable} term. Rather, named variables should be preferred when contents matter.

The correct way of formulating the original query is by explicitly stating the fact that all registers contain unusable and unrelated initial values, as shown below with eight different atoms.

\exam
\verb"s([xmm0, xmm1, xmm2, xmm3, xmm4, xmm5, xmm6, xmm7], I," \\
\verb"  [[0,0,0,0] | _ ])."
\maxe

This query will prompt the following list as a first solution, indicating a single instruction way of zeroing out register \texttt{xmm0}.

\exam
\verb"I = [i(pxor,xmm0,xmm0)]"
\maxe

\subsection{Iterative Deepening Search}

Prolog's DFS (depth-first search) is not very suited for this particular kind of state space search problem, since it will continuously append instructions to existing partial solutions in an attempt to reach the goal state. A BFS (breadth-first search) works much better, since it will report the \emph{shortest} instruction sequences from the initial state to the goal state first. We will implement such a search using Prolog's DFS, but without the inherently high memory demands of BFS, using IDS (iterative deepening search). To this end, the search rules given earlier are refined into the following set.
\exam
\verb"s(S, I, T) :- count(D, 0), s(S, I, T, D)." \\
\verb"s(S, [], S, 0 )." \\
\verb"s(S, [i(I,R1,R2)|J], T, X) :- X > 0, Y is X - 1," \\
\verb"                              i(I, R1, R2, S, U), s(U, J, T, Y)." \\
\maxe
The search rules themselves are as before, but restricted to a given depth. The \texttt{count} rules define a simple increment mechanism.
\exam
\verb"count(X, X)." \\
\verb"count(X, Y) :- Z is Y + 1, count(X, Z)."
\maxe

Combined, these rules try to find solutions within subsequent instruction sequences of length 0, 1, 2, etc. As a result, shorter instruction sequences are reported first (note that with some effort, this search mechanism can be adapted for other criteria of the \emph{best} solution, such as finding the instruction sequences with the lowest total cycle counts). For example, running the query
\exam
\verb"s( [ xmm0, xmm1, xmm2, xmm3, xmm4, xmm5, xmm6, xmm7], I," \\
\verb"   [ [-1,-1,-1,-1] | _ ] )." \\
\maxe
reports the desired solution
\exam
\verb"I = [i(pcmpeqd,xmm0,xmm0)]"
\maxe
before it reports the following alternative, but longer solution, which basically just clobbers the register with an unused value before resorting to the shorter solution.
\exam
\verb"I = [i(pxor,xmm0,xmm0),i(pcmpeqd,xmm0,xmm0)]"
\maxe

Suppose we are interested in broadcasting a value to all elements in a SIMD register, an idiom that is frequently used in the prologue of a vector loop by a vectorizing compiler~\cite{bikbook}. An instruction sequence for such a broadcast can be found using the following query, where atom \texttt{c} denotes the value that needs broadcasting.
\exam
\verb"s( [ xmm0, xmm1, xmm2, xmm3, xmm4, xmm5, xmm6, xmm7], I," \\
\verb"   [ [c,c,c,c] | _ ] )."
\maxe
Lacking a shuffle operation in our simplified rule set, the shortest instruction sequence for the broadcast consists of a data movement instruction followed by two unpack instructions.
\exam
\verb"I = [" \\
\verb"     i(movd,xmm0,c)," \\
\verb"     i(punpckldq,xmm0,xmm0)," \\
\verb"     i(punpckldq,xmm0,xmm0)" \\
\verb"    ]"
\maxe

\subsection{Usable Start State}

The examples so far searched for a particular goal state given an \emph{unusable} state state. Often, however, the start state may contain some known, usable information. As a simple example, the query
\exam
\verb"s([xmm0, [1,2,3,4], xmm2, xmm3, xmm4, xmm5, xmm6, xmm7], I," \\
\verb"  [[1,2,3,4] | _ ])."
\maxe
yields the following solution, which indicates that the best way to assign particular contents to register \texttt{xmm0} given a state where register \texttt{xmm1} already has these contents is simply moving the register. 
\exam
\verb"I = [i(movdqa,xmm0,xmm1)]"
\maxe

As a more practical application, this approach can be used to find the best sequence to sum up all elements in a SIMD register "horizontally", an idiom used by a vectorizing compiler~\cite{bikbook} to finalize the computation after converting a sum-reduction loop into SIMD code. Here the query
\exam
\verb"s( [ [a,b,c,d], xmm1, xmm2, xmm3, xmm4, xmm5, xmm6, xmm7], I" \\
\verb"   [ [_,_,_,(d+b)+(c+a)] | _ ] )."
\maxe
yields the following instruction sequence as first suitable solution.
\exam
\verb"I = [" \\
\verb"     i(movdqa,xmm1,xmm0)," \\
\verb"     i(psrldq,xmm0,8)," \\
\verb"     i(paddd,xmm1,xmm0)," \\
\verb"     i(punpckldq,xmm0,xmm1)," \\
\verb"     i(paddd,xmm0,xmm1)," \\
\verb"     i(psrldq,xmm0,4)" \\
\verb"    ]"
\maxe
Subsequent solutions with the same length provide some true alternatives (here, cycle counts could help finding the truly best one).
\exam
\verb"I = [" \\
\verb"     i(movdqa,xmm1,xmm0)," \\
\verb"     i(psrldq,xmm0,8)," \\
\verb"     i(paddd,xmm1,xmm0)," \\
\verb"     i(movdqa,xmm0,xmm1)," \\
\verb"     i(psrldq,xmm1,4)," \\
\verb"     i(paddd,xmm0,xmm1)" \\
\verb"    ]"
\maxe
Other solutions of the same length that follow may simply provide the same instruction sequences using different intermediate registers.

Note that in this example, a \emph{statically} known property of the context in which the instruction sequence is needed allowed for adding some usable information to the start state (viz. the SIMD register contains four partial results that need to be summed up). As stated in the introduction, at runtime the compiler could even exploit some \emph{dynamically} known properties of the context to find better instruction sequences, but it is unlikely that exhaustive search (let alone Prolog) could be used under such circumstances.

\section{Conclusions}

In this note, we explored using Prolog for finding efficient data manipulating instruction sequences. The problem is expressed as a state search problem, with the initial memory and register contents as start state, machine instructions as transitions, and the desired memory and register contents as goal state. Modeling instructions with a complete and accurate Prolog rule set of transitions will yield the best results, where it preferable to extract such a rule set automatically from an instruction set description in an electronic format. Search is expressed with Prolog rules as well, enhanced with iterative deepening to work around obvious complications with the default depth-first search of Prolog.

Once the best solution is found after exhaustively searching the state space, the instruction sequence can be hard-wired in any code generator that needs the data manipulation, and becomes at the immediate disposal of the compiler. Here, the best can be defined as the shortest instruction sequence or, with some adaptation, as the instruction sequence with minimal total weight, such as summing the cycle counts. For the sake of brevity, we restricted our focus on finding efficient Intel SSE instruction sequences using just a subset of the SIMD state, instructions, and data types. However, the presented ideas easily generalize to broader instructions sets and code generation problems.

\bibliography{bikbiblio}
\bibliographystyle{alpha}

\end{document}